\pdfoutput=1
\documentclass[reprint, aps, prd, preprintnumbers, amsmath, amssymb, superscriptaddress, nofootinbib]{revtex4-1}
\usepackage{amssymb,amsmath,bbold,slashed}
\usepackage{graphicx,soul}
\usepackage[usenames,dvipsnames]{xcolor}
\usepackage[unicode=true,pdfusetitle,
 bookmarks=true,bookmarksnumbered=false,bookmarksopen=false,citecolor=Turquoise,
 breaklinks=false,pdfborder={0 0 1},backref=false,colorlinks=true,pdfpagemode=FullScreen]
 {hyperref}

\usepackage{ulem}

\usepackage{graphicx}
\usepackage{dcolumn}
\usepackage{hyperref}

\begin{document}



\title{Probing Non-Thermal Gravitinos Through Large-Scale Structure Observations}

\author{Liyuan Guo}
\email{guoly25@mail2.sysu.edu.cn}
\affiliation{School of Physics, Sun Yat-Sen University, Guangzhou 510275, P. R. China}

\author{Weiyi Deng}
\email{dengwy23@mail2.sysu.edu.cn}
\affiliation{School of Physics, Sun Yat-Sen University, Guangzhou 510275, P. R. China}

\author{Chengcheng Han}
\email{hanchch@mail.sysu.edu.cn}
\affiliation{School of Physics, Sun Yat-Sen University, Guangzhou 510275, P. R. China}
\affiliation{Asia Pacific Center for Theoretical Physics, Pohang 37673, Korea}

\author{Zhanhong Lei}
\email{leizhh3@mail2.sysu.edu.cn}
\affiliation{School of Physics, Sun Yat-Sen University, Guangzhou 510275, P. R. China}

\date{\today}

\begin{abstract}
We investigate the effects of non-thermally produced dark matter on large-scale structure formation, focusing on the gravitino dark matter. Our analysis shows that large-scale structure measurements from the SDSS LRG data offer a promising approach to probing non-thermally produced gravitinos. Specifically, if gravitinos resulting from neutralino decay constitute a fraction  $F_\text{dec}=0.35~(0.1)$ of the dark matter component, the data exclude bino-like neutralino with a mass  $m_{\chi_1^0}\lesssim 25~(11)\text{ GeV}$ at $95\%\text{ C.L.}$. Furthermore, this constraint appears to be largely independent of the gravitino mass.

\end{abstract}

\maketitle

\section{\label{sec.1}Introduction}
One of the greatest mysteries in our universe is the existence of dark matter. Numerous astrophysical observations strongly suggest the presence of dark matter. It is estimated to constitute about one-quarter of the total energy budget of the universe. The quest to understand dark matter is crucial for explaining the formation and evolution of large-scale structures in the cosmos. Dark matter does not emit, absorb, or reflect light, making it invisible and detectable only through its gravitational effects. This enigmatic substance plays a vital role in the universe's structure, influencing the formation of galaxies, galaxy clusters, and other cosmic structures. Despite its elusive nature, understanding dark matter is essential for a comprehensive picture of the universe.

One of the popular dark matter candidates is the Weakly Interacting Massive Particle(WIMP). WIMPs are assumed to have a relic abundance resulting from the freeze-out process in the early universe. This process typically requires a moderate coupling with the standard model sector, making WIMPs feasible targets for both direct and indirect dark matter detection experiments. Among the WIMP candidates, the neutralino, which is usually hypothesized to be the lightest supersymmetric particle within the framework of supersymmetry(SUSY) ~\cite{Golfand:1971iwegdsf,Volkov1973ykgwj,Wess1974sbnny,Salam1974jyfg,Wess1974adtukf,Ferrara1974wkxfdk}, provide an interesting possibility~\cite{Steigman1985werfd, Jungman1996hrte, MARTIN1998wghfd, Feng2010fjshj, Cao:2012rz, Cao:2013gba, Han:2013usa, Han:2013kga, Han:2015lma, Han:2016xet, Han:2016gvr}. The minimal supersymmetric standard model(MSSM) is a well-motivated theoretical framework that not only provides a candidate for dark matter but also addresses other significant issues in particle physics, such as the naturalness problem and the unification of gauge couplings.

However, the search for dark matter signals through direct detection experiments, as well as the search for new particles at colliders~\cite{ATL-PHYS-PUB-2023-005}, does not find the evidence of the SUSY particles~\cite{Baer:2020kwz,Wang:2022rfd,Yang:2022qyz}. The non-observation of dark matter signals and the lack of evidence for supersymmetric particles in these experiments have led some researchers to consider alternative candidate of dark matter instead of neutralino~\cite{Ellis:2003dn,Covi:2001nw}. 


One alternative dark matter candidate is the super-WIMP~\cite{Feng:2003uy,Feng:2003xh, Feng:2004mt}, which interacts even more weakly with standard model particles than traditional WIMPs. The relic abundance of super-WIMPs is derived from the decay of a heavier particle that froze out in the early universe but decayed later due to its weak interactions with the super-WIMP. A natural candidate for a super-WIMP is the gravitino. Since a natural SUSY scale should be not much beyond TeV scale, the mass of the bino-like neutralino could vary from GeV to (few) TeV.
In scenarios where SUSY breaking occurs at a much lower energy scale than the Planck scale, the gravitino could have a much smaller mass(keV-GeV) than the neutralino. In this case, the neutralino first freezes out during the early universe, but later decays into the gravitino. The number density of gravitinos inherits from the neutralino.

If the neutralino is much heavier than the gravitino, the gravitino would acquire a large momentum at the moment of neutralino decaying, and streaming freely later on. If the free-streaming length is larger than a certain scale, it can lead to variations in the large-scale structure of the universe, therefore observations of the large-scale structure could provide feasible way to probe such super-WIMP candidates. Previous studies along this line can be found in~\cite{Nemevsek:2022anh, Nemevsek:2023yjl}  where a right-handed neutrino warm dark matter is considered.  Recent studies on other limits of gravitino dark matter can be found in~\cite{Deshpande:2023zed, Jodlowski:2023yne}.

In this work we focus on the effects of non-thermally produced gravitinos on the large-scale structure. The gravitinos can also be thermally produced through the freeze-in process during the early universe~\cite{Moroi:1993mb, Khlopov:1984pf, Hall:2009bx, Eberl:2020fml}, which is highly dependent on the reheating temperature. Given the current lack of understanding for UV physics, we treat the fraction of non-thermally produced gravitinos as a free parameter. By examining the impacts on large-scale structure, we provide insights into the viability of super-WIMP gravitinos as the dark matter candidates. The paper is organized as follows. In Sec.~\ref{sec.2} we briefly overview the physics relate to the gravitino dark matter. In Sec.~\ref{sec.3} we present our method to estimate the effect of non-thermally produced dark matter on the large scale structure and present our numerical results. We draw our conclusion in Sec.~\ref{sec.5}.

\section{\label{sec.2}Super-WIMP Gravitino}

In this section, we provide a brief overview of super-WIMP gravitino dark matter. Within the framework of SUSY, heavier SUSY particles undergo cascading decays into lighter ones, ultimately resulting in the lightest SUSY particle (LSP). If the gravitino is lighter than the neutralino, it becomes the LSP and a super-WIMP dark matter candidate, while the neutralino serves as the next-to-lightest SUSY particle (NLSP).
In the early universe, the neutralino would decay into the gravitino and Standard Model (SM) particles. If this decay occurs after Big Bang Nucleosynthesis (BBN), the resulting electromagnetic energy injection could alter the abundance of light elements~\cite{Feng:2003uy}. To avoid such effects, we consider only decays occurring before BBN, requiring the NLSP lifetime to be less than one second.
Since super-WIMPs interact extremely weakly with Standard Model particles, they never reach thermal equilibrium in the early universe. Instead, they can only be produced through the decay of heavier particles or via the thermal freeze-in process.

For a general super-WIMP scenario, the LSP inherits the abundance of the NLSP, and the LSP resulting from the NLSP may accounts for all dark matter. However, it is possible that not all LSPs are produced by NLSPs. Particularly, the gravitino could be produced by the thermal freeze-in in the early universe. The total super-WIMP abundance is then given by,
\begin{equation}
\Omega_\text{LSP}h^2 = \Omega_\text{LSP}^\text{thermal}h^2 + \Omega_\text{LSP}^\text{decay}h^2~,
\end{equation}
where $\Omega$ represents the relic abundance of the LSP, i.e., dark matter, and $h$ is the reduced Hubble parameter. Once the yield of NLSP, $Y_\text{NLSP} = n_\text{NLSP}/s$, is known, we can determine how much the decay-produced LSP contributes to the total dark matter abundance, where $n_\text{NLSP}$ is the number density of the NLSP and $s$ is the entropy density. Given that not all LSPs in this paper are decay-produced, we introduce
\begin{equation}
F_\text{dec} = \frac{\Omega_\text{LSP}^\text{decay}}{\Omega_{\text{DM}}}~,
\end{equation}
as the parameter to indicate the fraction of dark matter from the decay of the NLSP. For instance, $F_\text{dec} = 0.2$ implies that $20\%$ of the dark matter is decay-produced LSP. For more comprehensive discussion on super-WIMPs, one can refer to \cite{Cyburt:2002uv, Feng:2003xh, Rychkov:2007uq, Holtmann:1998gd}.

\subsection{\label{sec.2.1} Gravitinos from Neutralinos decay}

In our study, we focus on gravitinos, denoted by $\tilde{G}$, as potential super-WIMP dark matter candidate. Gravitinos are spin $3/2$ fermions and are the superpartners of gravitons. The gravitino mass depends on the SUSY breaking mechanism and is approximately given by $m_{\tilde{G}} \simeq \langle F \rangle/m_{\rm Pl}$, where $\langle F \rangle$ is the SUSY breaking scale and $m_{\rm Pl}$ is the Planck mass. Given that $\langle F \rangle$ is not tightly constrained, the gravitino mass is effectively a free parameter.

We explore the scenario where neutralinos decay into gravitinos via a two-body decay process: $\chi_1^0 \rightarrow \tilde{G} \gamma$. We consider only a bino-like neutralino $\chi^0_1$ decaying into a gravitino, as there are stringent constraints on wino- or higgsino-like neutralinos by colliders~\cite{ATLAS:2023meo,ATLAS:2018nud}. The decay width for a bino-like neutralino decaying into a gravitino is given by~\cite{Feng:2003uy},
\begin{equation}
\Gamma_{\chi_1^0}
=\frac{m_{\chi_1^0}^5\cos^2\theta_\text{W}}
{48\pi m_\text{Pl}^2m_{\tilde{G}}^2}
\left({
1-\frac{m_{\tilde{G}}^2}{m_{\chi_1^0}^2}
}\right)^3\left({
1+3\frac{m_{\tilde{G}}^2}{m_{\chi_1^0}^2}
}\right)~,
\label{eq.2.1.1}
\end{equation}
where $m_\text{Pl}$ is the reduced Planck mass, and $\theta_\text{W}$ is the weak mixing angle. After the decay of the neutralino, the photon’s energy, $E_{\gamma}$, would be re-distributed into the thermal plasma. Here we always assume the energy density of the radiation is not much affected by the photons from the decay of the neutralino. 
After the decay of the neutralinos, the gravitinos are still relativistic because of the large mass difference between the neutralino and gravitino, then the gravitinos are streaming freely in the early universe, suppressing the large scale structure formation. To estimate this effect, we should calculate the phase space distribution of the gravitinos after the decay of the neutralinos.

\subsection{\label{sec.2.2} Phase space distribution of the gravitinos}

\begin{figure}
\includegraphics[width=0.46\textwidth]{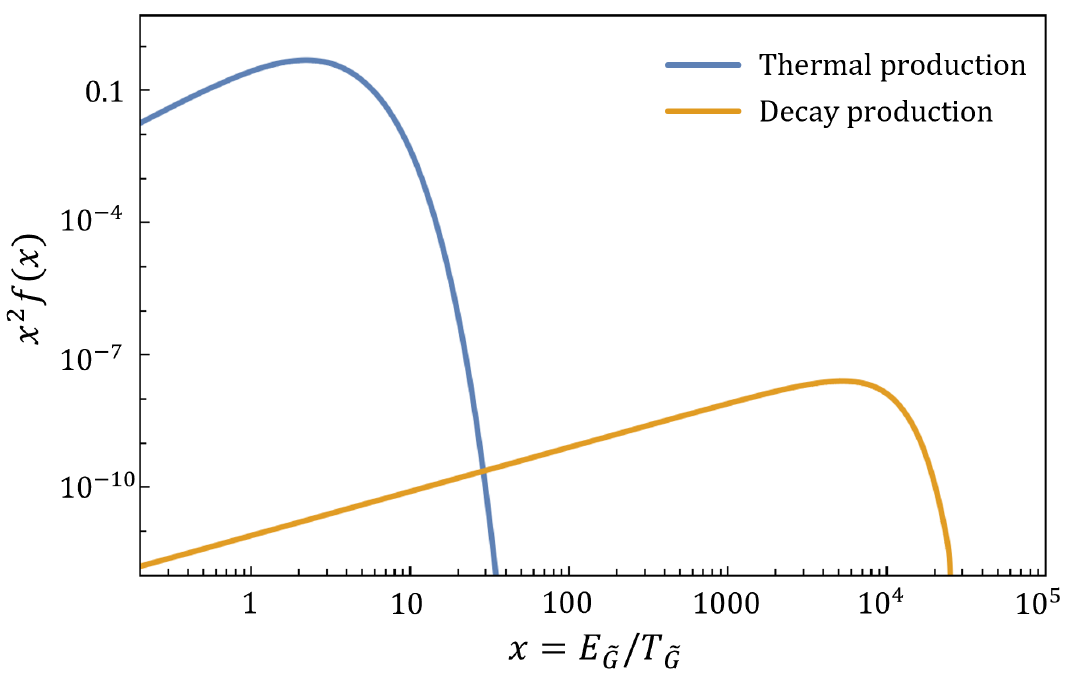}
\caption{Phase space distribution of gravitinos. The blue curve represents the Fermi-Dirac distribution. The orange curve illustrates the decayed gravitino model, with parameters $m_{\chi_1^0}=50\text{ TeV}, m_{\tilde{G}}=100\text{ keV},F_\text{dec}=0.1$.} \label{fig.1} 
\end{figure}

To calculate the phase space distribution of decay-produced gravitinos $f_{\tilde{G}}$, we can write down the evolution equation for the gravitino,
\begin{equation}
\frac{\partial f_{\tilde{G}}}{\partial t}-Hp\frac{\partial f_{\tilde{G}}}{\partial p}=C_\text{dec}~,
\label{eq.2.2.1}
\end{equation}
where $C_\text{dec}$ is the decay term. Assuming that the neutralino mass is much larger than the gravitino mass, it can be expressed as (see Appendix~\ref{app.1}),
\begin{equation}
C_\text{dec}=\frac{2\pi^2}{E_{\tilde{G}}^2}
\frac{\rho_{\chi_1^0}\Gamma_{\chi_1^0}}{g_{\chi_1^0}m_{\chi_1^0}}
\delta_\text{D}^{(1)}\left({E_{\tilde{G}}-\frac{1}{2}m_{\chi_1^0}}\right)~,
\end{equation}
where $g_{\chi_1^0}$ is the degeneracy factor of neutralinos.
The second term on the left of Eq.~\ref{eq.2.2.1} describes the contribution of the Hubble expansion. Introducing a Hubble expansion-independent variable $x \equiv E_{\tilde{G}}/T_{\tilde{G}}$, where $T_{\tilde{G}} \propto a^{-1}$ is a characteristic momentum of the gravitino which is set to be thermal temperature around the time of the neutralino decay. The Boltzmann equation for gravitinos can be expressed as follows (see Appendix~\ref{app.1}),
\begin{equation}
\frac{\partial f_{\tilde{G}}(x,t)}{\partial t}
=\frac{4\pi^2}{x^2}
\frac{\Gamma_{\chi_1^0}}{g_{\chi_1^0}m_{\chi_1^0}^2}
\frac{\rho_{\chi_1^0}(t)}{T_{\tilde{G}}(t)}
\delta_\text{D}^{(1)}\left[{T_{\tilde{G}}(t)-\frac{1}{2}\frac{m_{\chi_1^0}}{x}}\right]~.
\label{eq.2.2.3}
\end{equation}
Then the time integral over Eq.~\ref{eq.2.2.3} gives the phase space distribution of decay-produced gravitinos. For convenience, we introduce a new time variable $\zeta \equiv \ln a$ and the integral can be expressed as
\begin{align}
f_{\tilde{G}}(x)&=\frac{4\pi^2}{x^2}
\frac{\Gamma_{\chi_1^0}}{g_{\chi_1^0}m_{\chi_1^0}^2}
\int_{\zeta_\text{ini}}^{\zeta_\text{end}}\mathrm{d}\zeta
\frac{\rho_{\chi_1^0}(\zeta)}{T_{\tilde{G}}(\zeta)H(\zeta)} \nonumber \\
&\qquad\qquad\qquad\qquad
\times\delta_\text{D}^{(1)}\left[{T_{\tilde{G}}(\zeta)-\frac{1}{2}\frac{m_{\chi_1^0}}{x}}\right] \\
&=\frac{4\pi^2}{x^2}\frac{\Gamma_{\chi_1^0}}{g_{\chi_1^0}m_{\chi_1^0}^2}
\frac{\rho_{\chi_1^0}(\zeta_*)}{H(\zeta_*)T_{\tilde{G}}^2(\zeta_*)}~,
\label{eq.2.2.4}
\end{align}
where
\begin{equation}
\zeta_* \equiv \ln\left[{2x\frac{T_{\tilde{G}}(\zeta_\text{ini})}{m_{\chi_1^0}}}\right]~,
\end{equation}
identifies the exact moment at which the decay occurs. The initial time of integration is taken when $T = m_{\chi_1^0}/10$, and the ending moment is taken when $t = 10\ \tau_{\chi_1^0}$. The result is illustrated in Fig.~\ref{fig.1}. As comparison, we also show the thermal equilibrium distribution of the gravitino produced from the thermal freeze-in process with a Fermi-Dirac distribution,
\begin{equation}
f_\text{FD}(x)=\frac{1}{\mathrm{e}^x+1},\quad x\equiv\frac{E_{\tilde{G}}}{T_{\tilde{G}}}~.
\end{equation}
It clearly shows that the decay-produced gravitinos take much large energy than the thermally produced ones.

It is also necessary to determine how the energy density of neutralinos and the Hubble rate evolve with time. The corresponding coupled Boltzmann equations and Friedmann equation are,
\begin{align}
\dot{\rho}_{\chi_1^0} + 3H\rho_{\chi_1^0}
&= -\Gamma_{\chi_1^0}\rho_{\chi_1^0}~, \\
\dot{\rho}_{\tilde{G}} + 4H\rho_{\tilde{G}}
&= \frac{1}{2}\Gamma_{\chi_1^0}\rho_{\chi_1^0}~, \\
\dot{\rho}_{SM} + 4H\rho_{SM}
&= \frac{1}{2}\Gamma_{\chi_1^0}\rho_{\chi_1^0}~, \\
H^2 &= \frac{8\pi G}{3}
\left( \rho_{\chi_1^0} + \rho_{\tilde{G}} + \rho_{SM} \right)~.
\end{align}

Numerically solving the above equations yields the time dependence of the energy density of neutralinos and the Hubble rate. The initial time of integration is taken when $T = m_{\chi_1^0}/10$, the ending moment is taken when $t = 10\ \tau_{\chi_1^0}$ and the initial number density of decay-produced gravitinos vanishes. Since the number density of gravitinos inherits from the neutralino, we can determine the initial number density of neutralinos by ensuring that the number densities of gravitinos are consistent with present observations,
\begin{equation}
\frac{n_{\tilde{G}}(\zeta_0)}{s(\zeta_0)}
= \frac{n_{\tilde{G}}(\zeta_\text{end})}{s(\zeta_\text{end})}~,
\end{equation}
where $s$ represents the entropy density and $\zeta_0$ represent the scale factor of current time.

\section{\label{sec.3} Constraint from large scale structure}

We assume that the gravitinos are produced both by thermal freeze-in and neutralinos decay, and the fraction of decay-produced gravitinos is denoted by $F_\text{dec} \equiv {\Omega_\text{LSP}^\text{decay}}/{\Omega_{\text{DM}}}$. Due to the free streaming of the gravitino, they could alter the large scale structure. To estimate this effect, we utilize the \texttt{CLASS} (the Cosmic Linear Anisotropy Solving System) \cite{lesgourgues2011cosmic, Blas:2011, Lesgourgues:2011} to simulate the evolution of perturbations in the universe and obtain the present three-dimensional matter power spectrum. We use the LRG observations of large-scale structures or Lyman-$\alpha$ data to examine the parameter space of gravitino mass versus neutralino mass.

\subsection{\label{sec3.1}Matter Power Spectrum}

Since the gravitinos from the decay process inherit large energy from the neutralinos, if there is a large mass hierarchy between the neutralino and gravitino, the produced gravitinos will be relativistic and stream freely in the early universe. It will suppress the formation of large cosmic structure, which can be observed in the power spectrum. Additionally, if the thermally produced gravitinos have a mass as low as keV, they will also have a non-negligible velocity after decoupling. In this case gravitino could be a warm dark matter candidate, which may also suppress the growth of large scale structure. Lyman alpha forest observation already set a limit around $5\text{ keV}$ on the gravitino mass. Here we focus on the case where the mass of the gravitino heavier than $5\text{ keV}$ and treat the thermally produced gravitino as part of ``cold" dark matter.

For the thermally produced gravitinos, we assume they follows a Fermi-Dirac distribution. The gravitinos from the decay of neutralino are more energetic, as shown in Fig.~\ref{fig.1}. Particularly the decay-produced gravitinos generally does not follow Fermi-Dirac distribution, and we can take the typical physical momentum as their temperature.

\begin{figure}[t]
\includegraphics[width=0.46\textwidth]{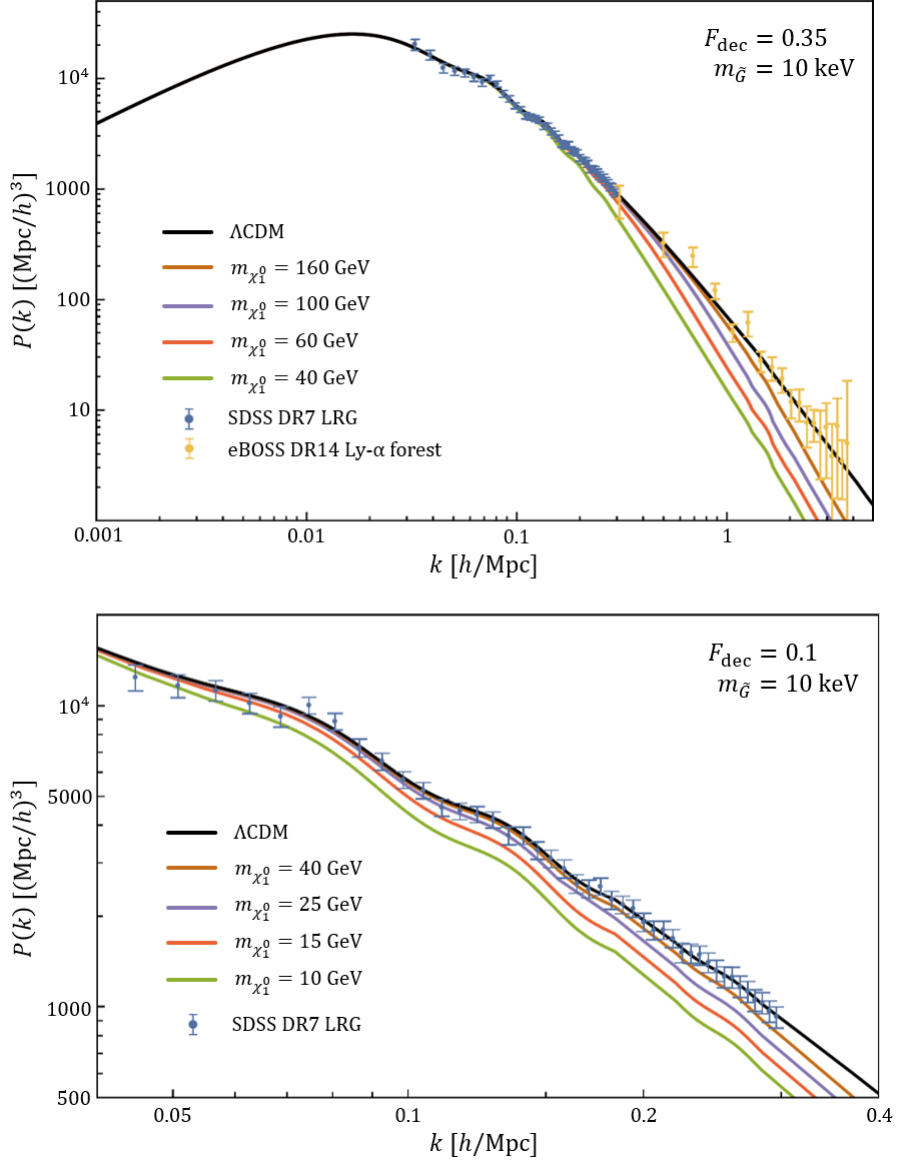}
\caption{\underline{Upper panel:} Linear matter power spectrum in standard $\Lambda$CDM (black solid curve) and decayed gravitino models with various neutralino mass (colorful solid curves). Here we set the gravitino mass $m_{\tilde{G}}=10\text{ keV}$ and the decayed gravitino fraction $F_\text{dec}=0.35$. Data points from SDSS DR7 LRG and Lyman-$\alpha$ observations are shown in blue and orange, respectively. \underline{Lower panel:} Same as the upper panel but for $F_\text{dec}=0.1$, with a different range for the power spectrum and various neutralino masses. Lyman-$\alpha$ data are not shown in this panel.} \label{fig.2} 
\end{figure}

In Fig.~\ref{fig.2}, we show the predicted linear matter power spectrum obtained for different neutralino masses. 
We find that lighter neutralinos lead to a stronger suppression on the power spectrum. This occurs because, as shown in Eq.~\ref{eq.2.1.1}, lighter neutralinos decay later. With less Hubble redshift, the decay-produced gravitinos travel faster, thus reducing more matter perturbations. The gravitino mass is fixed to be $10\text{ keV}$ as it does not affect the results, which will be explained in more detail later.
The nonlinear power spectrum of matter is calculated by the \texttt{HALOFIT}~\cite{Smith:2002dz}. 



\subsection{\label{sec3.2}Constraints from LRG Halo Power Spectrum}

Reid et al. \cite{reid2010cosmological} have provided the measured halo power spectrum for $0.02 < k < 0.2\ h\text{ Mpc}^{-1}$ from a sample of luminous red galaxies from the SDSS DR7. To compare with the predicted halo power spectrum, we need transform the predicted matter power spectrum into the halo power spectrum. We adopt the model given by Reid et al. \cite{reid2010cosmological},
\begin{equation}
P_\text{halo}(k) = P_\text{damp}(k)\,r_{\text{DM,damp}}(k)\,r_{\text{halo,DM}}(k)\,F_\text{nuis}(k)~,
\end{equation}
where $P_\text{damp}$ denotes the damped linear power spectrum, characterizing the damping of BAO and $r_\text{DM,damp}$ converts the damped linear power spectrum into the real-space non-linear matter power spectrum. $r_\text{halo,DM}$ transforms the matter power spectrum into the halo power spectrum, and $F_\text{nuis}$ accounts for smooth deviations from the model arising from modeling uncertainties.

Using the measured halo power spectrum, we can calculate the likelihood for our dark matter model,
\begin{equation}
\chi^2 = \sum \Delta_i C_{ij}^{-1} \Delta_j~,
\end{equation}
where $\Delta_i \equiv \hat{P}_\text{halo}(k_i) - P_\text{halo}(k_i)$. $\hat{P}_\text{halo}$ is the measured power spectrum of the reconstructed halo density field and $P_\text{halo}(k_i)$ is the theoretical predicted halo power spectrum for our model. We set the $\chi^2 > 60.4~(p < 0.05)$  as the $95\%$ limit in our model for 44 degree of freedom.


\subsection{\label{sec3.3}Constraints from  Lyman-$\alpha$ forest}

The production of gravitinos via neutralino decay is accompanied by a large momentum $p_\star\equiv(m_{\chi_1^0}^2-m_{\tilde{G}}^2)/(2m_{\chi_1^0})$ at the time of neutralino decay. Then this momentum redshifts as $p_{\tilde{G}}(t)=p_\star a_\text{prod}/a(t)$ due to expansion of the universe. The corresponding gravitino velocity is given by
\begin{align}
\begin{aligned}
v_{\tilde{G}}(t)
&=\frac{p_\star}{\sqrt{p_\star^2+m_{\tilde{G}}^2\,a^2(t)/a_\text{prod}^2}} \\
&=\frac{1}{\sqrt{1+a^2(t)/v_0^2}}~,
\end{aligned}
\end{align}
where
\begin{equation}
v_0\equiv\frac{p_\star}{m_{\tilde{G}}}a_\text{prod}~,
\end{equation}
is the present($a_0=1$) velocity of the  gravitino. This implies that gravitinos produced from neutralino decay will behave like warm dark matters(WDM), indicating that small-scale fluctuation measurements, such as the Lyman-$\alpha$ forest, can also impose constraints on the parameter space.

On scales smaller than a certain threshold, known as the free-streaming horizon, WDM significantly suppresses the growth of structures. On larger scales, however, it behaves similarly to cold dark matter. Given the time of production $t_\text{prod}$ and observation $t_\text{obs}$, the free-streaming horizon of gravitinos can be expressed as
\begin{align}
\lambda_\text{fs}&=\int_{t_\text{prod}}^{t_\text{obs}}
\frac{\mathrm{d}t}{a(t)}v_{\tilde{G}}(t)
=\int_{a_\text{prod}}^{a_\text{obs}}
\frac{\mathrm{d}a}{a^2H(a)}v_{\tilde{G}}(a)
\label{eq.3.3.3} \\
&=\frac{1}{H_0}\sqrt{\frac{a_\text{eq}}{\Omega_\text{m}}}
\int_{a_\text{dec}/a_\text{eq}}^{a_\text{obs}/a_\text{eq}}
\frac{\mathrm{d}y}{\sqrt{(1+y)[1+(a_\text{eq}/v_0)^2y^2]}}~,
\label{eq.3.3.4}
\end{align}
where $a_\texttt{eq}$ is the scale factor at matter-radiation equality, $\Omega_\text{m}$ is the present reduced total matter density, $H\equiv\dot{a}/a$ is the Hubble expansion rate, and $H_0$ is present Hubble expansion rate.

Assuming that all the dark matter of the Universe is constituted by the thermally produced WDM, the typical Lyman-$\alpha$ forest observations set a lower mass limit on thermally produced WDM, requiring $m_\text{WDM}\gtrsim 5\text{ keV}$ at $95\%$ confidence level (C.L.)~\cite{PhysRevD.96.023522}. The relationship between the temperature and mass of thermal WDM is given by
\begin{equation}
\Omega_\text{WDM}h^2
=\left(\frac{T_\text{WDM}}{T_\nu}\right)^3
\left(\frac{m_\text{WDM}}{94\text{ eV}}\right)~,
\end{equation}
where $T_\nu$ is the temperature of the neutrino background. Therefore the upper limit on the present velocity of WDM can be derived as $v_{\text{WDM,0}}\simeq3\,T_{\text{WDM},0}/m_\text{WDM}\lesssim1.2\times10^{-8}$. Based on the description in Eq.~\ref{eq.3.3.4} and letting $a_\text{prod}\to0$, this result can be converted into the free-streaming horizon constraint $\lambda_\text{fs}\lesssim0.045\,h^{-1}\text{Mpc}$ at $z_\text{obs}=2$.

Considering a scenario of mixed warm dark matter and cold dark matter,  Boyarsky et al. \cite{Boyarsky:2008xj} performs an analysis and presents a limit on the fraction of warm dark matter 
\begin{equation}
f_\text{WDM}\equiv\Omega_\text{WDM}/\Omega_\text{DM}~,
\end{equation}
for different velocities of the warm dark matter $v_0$.
We translated these into constraints on the free-streaming horizon, as shown in Table~\ref{tab.1}. For $f_\text{WDM}=1$,  \cite{Boyarsky:2008xj} show that an upper limit of $0.183\,h^{-1}\text{Mpc}$ for $\lambda_\text{fs}$, contrasting with more recent 2017 results \cite{PhysRevD.96.023522} of $\lambda_\text{fs}\lesssim 0.045\,h^{-1}\text{Mpc}$. For $f_\text{WDM}\lesssim 0.35$, there is no constraint on the warm dark matter \cite{Boyarsky:2008xj}. 

Using the constraint from free-streaming horizon, we also calculated the limitation on the gravitino mass in the scenario where $100\%$ of gravitino is thermally produced. We obtained a lower mass limit of $1.7\text{ keV}$ for gravitino, which is consistent with the results~\cite{Boyarsky:2008xj}.

\begin{table}[t]
\caption{\label{tab.1}Upper limit on the free-streaming horizon $\lambda_\text{fs}$ as a function of the WDM fraction $f_\text{WDM}$. These constraints are derived from the two-dimensional $95\%$ credible contours in the $(f_\text{WDM},v_{\text{WDM},0})$-plane, as presented in Fig.~12 of Ref.~\cite{Boyarsky:2008xj}.}
\begin{ruledtabular}
\begin{tabular}{cc}
$f_\text{WDM}$ & Upper limit on $\lambda_\text{fs}\ [h^{-1}\text{Mpc}]$ \\
\colrule
0.35 & 0.330  \\
0.4 & 0.294 \\
0.5 & 0.245 \\
0.6 & 0.221 \\
0.7 & 0.209 \\
0.8 & 0.191 \\
0.9 & 0.185 \\
1.0 & 0.183
\end{tabular}
\end{ruledtabular}
\end{table}

\subsection{\label{subsec.4} Numerical results}

\begin{figure}[h]
\includegraphics[width=0.46\textwidth]{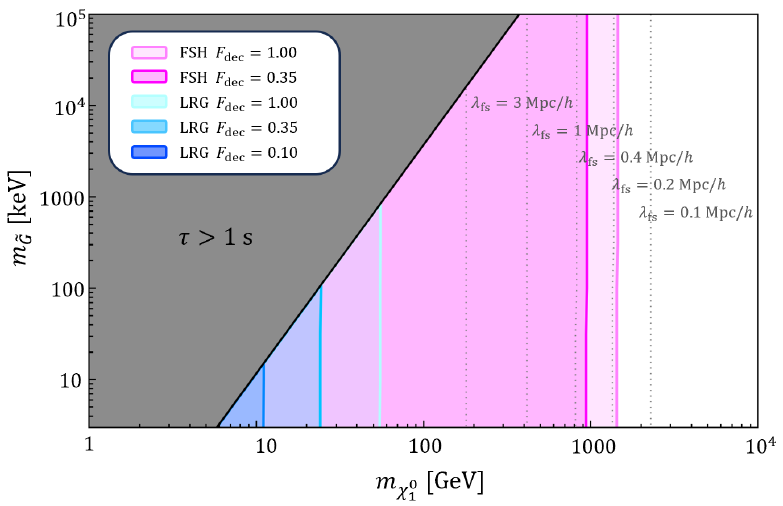}
\caption{LRG and Lyman-$\alpha$ constraints on the parameter space in the ($m_{\chi_1^0} , m_{\tilde{G}}$)-plane, depicted by blue and magenta solid curves, respectively. The colored regions are excluded. The gray region corresponds to a neutralino lifetime longer than BBN. The different free-streaming horizons are also illustrated with grey dotted lines.} \label{fig.3} 
\end{figure}

Varying the neutralino masses, gravitino masses and decay-produced gravitino fractions $F_\text{dec}$, we calculated the corresponding gravitino phase space distribution functions and obtained the matter power spectra, which are transformed into halo power spectra. Then we compared the predicted halo power spectra with the SDSS DR7 LRG measurements, and calculated the corresponding $p$-value and we set $p < 0.05$ as the  $95\%\text{ C.L.}$ limit. As shown in Fig.~\ref{fig.3}, the blue solid line indicates the LRG $95\%\text{ C.L.}$ limit, which corresponds to $F_\text{dec}=0.1, 0.35, 1$ from left to right. The region marked ``$\tau>1\,\text{s}$" corresponds to the neutralino decay later than BBN.

As a comparison, we have also plotted the free-streaming horizon constraints transformed from the Lyman-$\alpha$ data, represented by the solid magenta line, which corresponds from left to right to $F_\text{dec}=0.35, 1$. Since the thermally produced gravitino is cold and the decay-produced gravitino behaves like WDM, the fraction $F_\text{dec}$ corresponds to the WDM fraction $f_\text{WDM}$ in Lyman-$\alpha$ constraints. The different free-streaming horizons are also drawn as references with gray dotted lines. Note that when $F_\text{dec}<0.35$, the constraints from Lyman-$\alpha$ forests are not available at $95\%\text{ C.L.}$~\cite{Boyarsky:2008xj}, and the constraints from the large scale structure become important. In addition, if the C.L. is relaxed to $68\%$, Lyman-$\alpha$ forests can still provide constraints for $F_\text{dec}>0.15$~\cite{Boyarsky:2008xj}.

When $F_\text{dec}=0.1$, the LRG measurements exclude gravitinos obtained from the decay of neutralinos with mass $m_{\chi_1^0}\lesssim 11\text{ GeV}$ at $95\%\text{ C.L.}$, and the dependence of the exclusion boundaries are weakly dependent on the gravitino mass. This is because the scale factor at gravitino production is nearly proportional to the gravitino mass, $a_\text{prod}\propto\Gamma_{\chi_1^0}^{-\frac{1}{2}}\propto m_{\tilde{G}}$, making the velocity $v_0\equiv\frac{p_\star}{m_{\tilde{G}}}a_\text{prod}$ independent of gravitino mass. Note that in Fig.~2 of Ref.~\cite{Deshpande:2023zed}, the Lyman-$\alpha$ constraints do not extend below $m_{\tilde{G}}\sim10^5\text{ keV}$, because they take the WDM fraction to be $f_\text{WDM}=m_{\tilde{G}}/m_{\chi_1^0}$. The fraction of WDM is highly suppressed when $m_{\tilde{G}} < 10^5\text{ keV}$.   

\section{\label{sec.5}Conclusions}

In this paper, we explore the impact of non-thermally produced gravitino dark matter on large-scale structure formation. We consider a scenario where bino-like neutralinos freeze out and subsequently decay into gravitinos. The significant mass difference between the neutralino and the gravitino imparts a high momentum to the gravitino, causing it to stream freely and suppress the growth of large-scale structures. Using large-scale structure measurements from the SDSS LRG data, we demonstrate that these observations offer a viable method to probe non-thermally produced gravitinos. For instance, if gravitinos produced from decay constitute a fraction of  $F_\text{dec}=0.35~(0.1)$ of the dark matter component, the large-scale structure data excludes bino-like neutralinos with masses  $m_{\chi_1^0}\lesssim 25~(11)\text{ GeV}$ at $95\%\text{ C.L.}$. We also find that this limit is nearly independent of the gravitino mass.


\begin{acknowledgments}
We acknowledge Yue Zhang for his helpful discussions. This work was supported by National Key R\&D Program of China under grant Nos. 2023YFA1606100. C. H. acknowledges supports from the National Natural Science Foundation of China (NSFC) under grant No 12435005, the Fundamental Research Funds for the Central Universities at Sun Yat-sen University under Grant No. 24qnpy117, and the Key Laboratory of Particle Astrophysics and Cosmology (MOE) of Shanghai Jiao Tong University.
\end{acknowledgments}

\appendix

\section{\label{app.1}Boltzmann Equation for Gravitinos}

Besides the freeze-in process during the early universe, the gravitino can also be produced by the decay of the bino-like neutralino, represented as
\begin{equation}
\chi_1^0\to\tilde{G}+\gamma~.
\end{equation}
The temporal evolution of the phase space distribution for the resulting gravitino is described by the Boltzmann equation, which connects the phase space distribution functions $f_{\chi_1^0}$ and $f_{\tilde{G}}$,
\begin{equation}
\frac{\partial f_{\tilde{G}}}{\partial t}-Hp\frac{\partial f_{\tilde{G}}}{\partial p}=C_\text{dec}~,
\label{eq.a.2}
\end{equation}
where $C_\text{dec}$ denotes the decay term. Assuming that the mass of the neutralino is significantly greater than that of the gravitino and working in the rest frame of $\chi_1^0$, the decay term can be expressed as
\begin{align}
\begin{aligned}
C_\text{dec}&=\frac{1}{2E_{\tilde{G}}}
\int\frac{\mathrm{d}^3p_{\chi_1^0}}{(2\pi)^3}\frac{1}{2m_{\chi_1^0}}
\int\frac{\mathrm{d}^3p_\gamma}{(2\pi)^3}\frac{1}{2E_\gamma} \\
&\qquad\times(2\pi)^4\delta_\text{D}^{(4)}(p_{\chi_1^0}-p_{\tilde{G}}-p_\gamma)
|\mathcal{M}|^2f_{\chi_1^0} \\
&=\frac{\pi}{4}\frac{1}{m_{\chi_1^0}E_{\tilde{G}}^2}
|\mathcal{M}|^2\delta_\text{D}^{(1)}(m_{\chi_1^0}-2E_{\tilde{G}})
\int\frac{\mathrm{d}^3p_{\chi_1^0}}{(2\pi)^3}f_{\chi_1^0} \\
&=\frac{\pi}{8}\frac{1}{E_{\tilde{G}}^2}
\frac{\rho_{\chi_1^0}}{g_{\chi_1^0}m_{\chi_1^0}^2}
|\mathcal{M}|^2\delta_\text{D}^{(1)}(E_{\tilde{G}}-\frac{1}{2}m_{\chi_1^0})~,
\end{aligned}
\end{align}
where $g_{\chi_1^0}$ represents the degeneracy factor of the neutralino and $\mathcal{M}$ denotes the decay matrix element associated with the neutralino decay. The decay width of the neutralino can be described as
\begin{align}
\begin{aligned}
\Gamma_{\chi_1^0}&=\frac{1}{2m_{\chi_1^0}}
\int\frac{\mathrm{d}^3p_{\tilde{G}}}{(2\pi)^3}\frac{1}{2E_{\tilde{G}}}
\int\frac{\mathrm{d}^3p_\gamma}{(2\pi)^3}\frac{1}{2E_\gamma} \\
&\qquad\times(2\pi)^4\delta_D^{(4)}(p_{\chi_1^0}-p_{\tilde{G}}-p_\gamma)|\mathcal{M}|^2 \\
&=\frac{1}{32\pi^2}\frac{1}{m_{\chi_1^0}}|\mathcal{M}|^2
\int\mathrm{d}\Omega\,\mathrm{d}p_{\tilde{G}}
\,\delta_D^{(1)}(m_{\chi_1^0}-2p_{\tilde{G}}) \\
&=\frac{1}{16\pi}\frac{1}{m_{\chi_1^0}}|\mathcal{M}|^2~.
\end{aligned}
\end{align}
Subsequently, we can relate $C_\text{dec}$ and $\Gamma_\text{dec}$ and simplify the Boltzmann equation to
\begin{equation}
C_\text{dec}=\frac{2\pi^2}{E_{\tilde{G}}^2}
\frac{\rho_{\chi_1^0}\Gamma_{\chi_1^0}}{g_{\chi_1^0}m_{\chi_1^0}}
\delta_\text{D}^{(1)}\left(E_{\tilde{G}}-\frac{1}{2}m_{\chi_1^0}\right)~.
\end{equation}

Next, for the left-hand side of Eq.~\ref{eq.a.2}, we introduce a variable $x\equiv E_{\tilde{G}}/T_{\tilde{G}}=p_{\tilde{G}}/T_{\tilde{G}}$, where $T_{\tilde{G}}\propto a^{-1}$ is a characteristic momentum of the gravitino which is set to be thermal temperature at around the time of the neutralino decay. This introduction is justified because as long as the gravitinos remain ultra-relativistic, the ratio $x$ stays invariant in the expanding universe. Using the identity
\begin{equation}
\frac{\partial f_{\tilde{G}}(p,t)}{\partial t}
-Hp\frac{\partial f_{\tilde{G}}(p,t)}{\partial p}
=\frac{\partial f_{\tilde{G}}(x,t)}{\partial t}~,
\end{equation}
we ultimately derive the phase space evolution equation for gravitinos resulting from neutralino decay,
\begin{align}
\begin{aligned}
\frac{\partial f_{\tilde{G}}(x,t)}{\partial t}
&=\frac{2\pi^2}{x^2T_{\tilde{G}}^2}
\frac{\rho_{\chi_1^0}\Gamma_{\chi_1^0}}{g_{\chi_1^0}m_{\chi_1^0}}
\delta_\text{D}^{(1)}\left(x\,T_{\tilde{G}}-\frac{1}{2}m_{\chi_1^0}\right) \\
&=\frac{2\pi^2}{x^3T_{\tilde{G}}^2}
\frac{\rho_{\chi_1^0}\Gamma_{\chi_1^0}}{g_{\chi_1^0}m_{\chi_1^0}}
\delta_\text{D}^{(1)}\left({T_{\tilde{G}}-\frac{1}{2}\frac{m_{\chi_1^0}}{x}}\right) \\
&=\frac{4\pi^2}{x^2}
\frac{\Gamma_{\chi_1^0}}{g_{\chi_1^0}m_{\chi_1^0}^2}
\frac{\rho_{\chi_1^0}}{T_{\tilde{G}}}
\delta_\text{D}^{(1)}\left({T_{\tilde{G}}-\frac{1}{2}\frac{m_{\chi_1^0}}{x}}\right)~.
\end{aligned}
\end{align}

\vspace{2em}

\bibliography{main}

\begin{thebibliography}{46}%
\makeatletter
\providecommand \@ifxundefined [1]{%
 \@ifx{#1\undefined}
}%
\providecommand \@ifnum [1]{%
 \ifnum #1\expandafter \@firstoftwo
 \else \expandafter \@secondoftwo
 \fi
}%
\providecommand \@ifx [1]{%
 \ifx #1\expandafter \@firstoftwo
 \else \expandafter \@secondoftwo
 \fi
}%
\providecommand \natexlab [1]{#1}%
\providecommand \enquote  [1]{``#1''}%
\providecommand \bibnamefont  [1]{#1}%
\providecommand \bibfnamefont [1]{#1}%
\providecommand \citenamefont [1]{#1}%
\providecommand \href@noop [0]{\@secondoftwo}%
\providecommand \href [0]{\begingroup \@sanitize@url \@href}%
\providecommand \@href[1]{\@@startlink{#1}\@@href}%
\providecommand \@@href[1]{\endgroup#1\@@endlink}%
\providecommand \@sanitize@url [0]{\catcode `\\12\catcode `\$12\catcode
  `\&12\catcode `\#12\catcode `\^12\catcode `\_12\catcode `\%12\relax}%
\providecommand \@@startlink[1]{}%
\providecommand \@@endlink[0]{}%
\providecommand \url  [0]{\begingroup\@sanitize@url \@url }%
\providecommand \@url [1]{\endgroup\@href {#1}{\urlprefix }}%
\providecommand \urlprefix  [0]{URL }%
\providecommand \Eprint [0]{\href }%
\providecommand \doibase [0]{http://dx.doi.org/}%
\providecommand \selectlanguage [0]{\@gobble}%
\providecommand \bibinfo  [0]{\@secondoftwo}%
\providecommand \bibfield  [0]{\@secondoftwo}%
\providecommand \translation [1]{[#1]}%
\providecommand \BibitemOpen [0]{}%
\providecommand \bibitemStop [0]{}%
\providecommand \bibitemNoStop [0]{.\EOS\space}%
\providecommand \EOS [0]{\spacefactor3000\relax}%
\providecommand \BibitemShut  [1]{\csname bibitem#1\endcsname}%
\let\auto@bib@innerbib\@empty
\bibitem [{\citenamefont {Golfand}\ and\ \citenamefont
  {Likhtman}(1971)}]{Golfand:1971iwegdsf}%
  \BibitemOpen
  \bibfield  {author} {\bibinfo {author} {\bibfnamefont {Y.~A.}\ \bibnamefont
  {Golfand}}\ and\ \bibinfo {author} {\bibfnamefont {E.~P.}\ \bibnamefont
  {Likhtman}},\ }\href {\doibase 10.1142/9789814542340_0001} {\bibfield
  {journal} {\bibinfo  {journal} {JETP Lett.}\ }\textbf {\bibinfo {volume}
  {13}},\ \bibinfo {pages} {323} (\bibinfo {year} {1971})}\BibitemShut
  {NoStop}%
\bibitem [{\citenamefont {Volkov}\ and\ \citenamefont
  {Akulov}(1973)}]{Volkov1973ykgwj}%
  \BibitemOpen
  \bibfield  {author} {\bibinfo {author} {\bibfnamefont {D.~V.}\ \bibnamefont
  {Volkov}}\ and\ \bibinfo {author} {\bibfnamefont {V.~P.}\ \bibnamefont
  {Akulov}},\ }\href {\doibase 10.1016/0370-2693(73)90490-5} {\bibfield
  {journal} {\bibinfo  {journal} {Physics Letters B}\ }\textbf {\bibinfo
  {volume} {46}},\ \bibinfo {pages} {109} (\bibinfo {year} {1973})}\BibitemShut
  {NoStop}%
\bibitem [{\citenamefont {Wess}\ and\ \citenamefont
  {Zumino}(1974{\natexlab{a}})}]{Wess1974sbnny}%
  \BibitemOpen
  \bibfield  {author} {\bibinfo {author} {\bibfnamefont {J.}~\bibnamefont
  {Wess}}\ and\ \bibinfo {author} {\bibfnamefont {B.}~\bibnamefont {Zumino}},\
  }\href {\doibase 10.1016/0550-3213(74)90355-1} {\bibfield  {journal}
  {\bibinfo  {journal} {Nuclear Physics B}\ }\textbf {\bibinfo {volume} {70}},\
  \bibinfo {pages} {39} (\bibinfo {year} {1974}{\natexlab{a}})}\BibitemShut
  {NoStop}%
\bibitem [{\citenamefont {Salam}\ and\ \citenamefont
  {Strathdee}(1974)}]{Salam1974jyfg}%
  \BibitemOpen
  \bibfield  {author} {\bibinfo {author} {\bibfnamefont {A.}~\bibnamefont
  {Salam}}\ and\ \bibinfo {author} {\bibfnamefont {J.}~\bibnamefont
  {Strathdee}},\ }\href {\doibase 10.1016/0370-2693(74)90226-3} {\bibfield
  {journal} {\bibinfo  {journal} {Physics Letters B}\ }\textbf {\bibinfo
  {volume} {51}},\ \bibinfo {pages} {353} (\bibinfo {year} {1974})}\BibitemShut
  {NoStop}%
\bibitem [{\citenamefont {Wess}\ and\ \citenamefont
  {Zumino}(1974{\natexlab{b}})}]{Wess1974adtukf}%
  \BibitemOpen
  \bibfield  {author} {\bibinfo {author} {\bibfnamefont {J.}~\bibnamefont
  {Wess}}\ and\ \bibinfo {author} {\bibfnamefont {B.}~\bibnamefont {Zumino}},\
  }\href {\doibase 10.1016/0550-3213(74)90112-6} {\bibfield  {journal}
  {\bibinfo  {journal} {Nuclear Physics, Section B}\ }\textbf {\bibinfo
  {volume} {78}},\ \bibinfo {pages} {1} (\bibinfo {year}
  {1974}{\natexlab{b}})}\BibitemShut {NoStop}%
\bibitem [{\citenamefont {Ferrara}\ and\ \citenamefont
  {Zumino}(1974)}]{Ferrara1974wkxfdk}%
  \BibitemOpen
  \bibfield  {author} {\bibinfo {author} {\bibfnamefont {S.}~\bibnamefont
  {Ferrara}}\ and\ \bibinfo {author} {\bibfnamefont {B.}~\bibnamefont
  {Zumino}},\ }\href {\doibase 10.1016/0550-3213(74)90559-8} {\bibfield
  {journal} {\bibinfo  {journal} {Nuclear Physics, Section B}\ }\textbf
  {\bibinfo {volume} {79}},\ \bibinfo {pages} {413} (\bibinfo {year}
  {1974})}\BibitemShut {NoStop}%
\bibitem [{\citenamefont {Steigman}\ and\ \citenamefont
  {Turner}(1985)}]{Steigman1985werfd}%
  \BibitemOpen
  \bibfield  {author} {\bibinfo {author} {\bibfnamefont {G.}~\bibnamefont
  {Steigman}}\ and\ \bibinfo {author} {\bibfnamefont {M.~S.}\ \bibnamefont
  {Turner}},\ }\href {\doibase 10.1016/0550-3213(85)90537-1} {\bibfield
  {journal} {\bibinfo  {journal} {Nuclear Physics B}\ }\textbf {\bibinfo
  {volume} {253}},\ \bibinfo {pages} {375} (\bibinfo {year}
  {1985})}\BibitemShut {NoStop}%
\bibitem [{\citenamefont {Jungman}\ \emph {et~al.}(1995)\citenamefont
  {Jungman}, \citenamefont {Kamionkowski},\ and\ \citenamefont
  {Griest}}]{Jungman1996hrte}%
  \BibitemOpen
  \bibfield  {author} {\bibinfo {author} {\bibfnamefont {G.}~\bibnamefont
  {Jungman}}, \bibinfo {author} {\bibfnamefont {M.}~\bibnamefont
  {Kamionkowski}}, \ and\ \bibinfo {author} {\bibfnamefont {K.}~\bibnamefont
  {Griest}},\ }\href {\doibase 10.1016/0370-1573(95)00058-5} {\bibfield
  {journal} {\bibinfo  {journal} {Physics Reports}\ }\textbf {\bibinfo {volume}
  {267}},\ \bibinfo {pages} {195} (\bibinfo {year} {1995})},\ \Eprint
  {http://arxiv.org/abs/9506380} {arXiv:9506380 [hep-ph]} \BibitemShut
  {NoStop}%
\bibitem [{\citenamefont {Martin}(1998)}]{MARTIN1998wghfd}%
  \BibitemOpen
  \bibfield  {author} {\bibinfo {author} {\bibfnamefont {S.~P.}\ \bibnamefont
  {Martin}},\ }\href {\doibase 10.1142/9789812839657_0001} {\bibfield
  {journal} {\bibinfo  {journal} {Adv. Ser. Direct. High Energy Phys.}\
  }\textbf {\bibinfo {volume} {18}},\ \bibinfo {pages} {1} (\bibinfo {year}
  {1998})},\ \Eprint {http://arxiv.org/abs/hep-ph/9709356}
  {arXiv:hep-ph/9709356} \BibitemShut {NoStop}%
\bibitem [{\citenamefont {Feng}(2010)}]{Feng2010fjshj}%
  \BibitemOpen
  \bibfield  {author} {\bibinfo {author} {\bibfnamefont {J.~L.}\ \bibnamefont
  {Feng}},\ }\href {\doibase 10.1146/annurev-astro-082708-101659} {\bibfield
  {journal} {\bibinfo  {journal} {Annual Review of Astronomy and Astrophysics}\
  }\textbf {\bibinfo {volume} {48}},\ \bibinfo {pages} {495} (\bibinfo {year}
  {2010})},\ \Eprint {http://arxiv.org/abs/1003.0904} {arXiv:1003.0904}
  \BibitemShut {NoStop}%
\bibitem [{\citenamefont {Cao}\ \emph {et~al.}(2012)\citenamefont {Cao},
  \citenamefont {Han}, \citenamefont {Wu}, \citenamefont {Yang},\ and\
  \citenamefont {Zhang}}]{Cao:2012rz}%
  \BibitemOpen
  \bibfield  {author} {\bibinfo {author} {\bibfnamefont {J.}~\bibnamefont
  {Cao}}, \bibinfo {author} {\bibfnamefont {C.}~\bibnamefont {Han}}, \bibinfo
  {author} {\bibfnamefont {L.}~\bibnamefont {Wu}}, \bibinfo {author}
  {\bibfnamefont {J.~M.}\ \bibnamefont {Yang}}, \ and\ \bibinfo {author}
  {\bibfnamefont {Y.}~\bibnamefont {Zhang}},\ }\href {\doibase
  10.1007/JHEP11(2012)039} {\bibfield  {journal} {\bibinfo  {journal} {JHEP}\
  }\textbf {\bibinfo {volume} {11}},\ \bibinfo {pages} {039} (\bibinfo {year}
  {2012})},\ \Eprint {http://arxiv.org/abs/1206.3865} {arXiv:1206.3865
  [hep-ph]} \BibitemShut {NoStop}%
\bibitem [{\citenamefont {Cao}\ \emph {et~al.}(2013)\citenamefont {Cao},
  \citenamefont {Ding}, \citenamefont {Han}, \citenamefont {Yang},\ and\
  \citenamefont {Zhu}}]{Cao:2013gba}%
  \BibitemOpen
  \bibfield  {author} {\bibinfo {author} {\bibfnamefont {J.}~\bibnamefont
  {Cao}}, \bibinfo {author} {\bibfnamefont {F.}~\bibnamefont {Ding}}, \bibinfo
  {author} {\bibfnamefont {C.}~\bibnamefont {Han}}, \bibinfo {author}
  {\bibfnamefont {J.~M.}\ \bibnamefont {Yang}}, \ and\ \bibinfo {author}
  {\bibfnamefont {J.}~\bibnamefont {Zhu}},\ }\href {\doibase
  10.1007/JHEP11(2013)018} {\bibfield  {journal} {\bibinfo  {journal} {JHEP}\
  }\textbf {\bibinfo {volume} {11}},\ \bibinfo {pages} {018} (\bibinfo {year}
  {2013})},\ \Eprint {http://arxiv.org/abs/1309.4939} {arXiv:1309.4939
  [hep-ph]} \BibitemShut {NoStop}%
\bibitem [{\citenamefont {Han}\ \emph {et~al.}(2014)\citenamefont {Han},
  \citenamefont {Kobakhidze}, \citenamefont {Liu}, \citenamefont {Saavedra},
  \citenamefont {Wu},\ and\ \citenamefont {Yang}}]{Han:2013usa}%
  \BibitemOpen
  \bibfield  {author} {\bibinfo {author} {\bibfnamefont {C.}~\bibnamefont
  {Han}}, \bibinfo {author} {\bibfnamefont {A.}~\bibnamefont {Kobakhidze}},
  \bibinfo {author} {\bibfnamefont {N.}~\bibnamefont {Liu}}, \bibinfo {author}
  {\bibfnamefont {A.}~\bibnamefont {Saavedra}}, \bibinfo {author}
  {\bibfnamefont {L.}~\bibnamefont {Wu}}, \ and\ \bibinfo {author}
  {\bibfnamefont {J.~M.}\ \bibnamefont {Yang}},\ }\href {\doibase
  10.1007/JHEP02(2014)049} {\bibfield  {journal} {\bibinfo  {journal} {JHEP}\
  }\textbf {\bibinfo {volume} {02}},\ \bibinfo {pages} {049} (\bibinfo {year}
  {2014})},\ \Eprint {http://arxiv.org/abs/1310.4274} {arXiv:1310.4274
  [hep-ph]} \BibitemShut {NoStop}%
\bibitem [{\citenamefont {Han}\ \emph {et~al.}(2013)\citenamefont {Han},
  \citenamefont {Hikasa}, \citenamefont {Wu}, \citenamefont {Yang},\ and\
  \citenamefont {Zhang}}]{Han:2013kga}%
  \BibitemOpen
  \bibfield  {author} {\bibinfo {author} {\bibfnamefont {C.}~\bibnamefont
  {Han}}, \bibinfo {author} {\bibfnamefont {K.-i.}\ \bibnamefont {Hikasa}},
  \bibinfo {author} {\bibfnamefont {L.}~\bibnamefont {Wu}}, \bibinfo {author}
  {\bibfnamefont {J.~M.}\ \bibnamefont {Yang}}, \ and\ \bibinfo {author}
  {\bibfnamefont {Y.}~\bibnamefont {Zhang}},\ }\href {\doibase
  10.1007/JHEP10(2013)216} {\bibfield  {journal} {\bibinfo  {journal} {JHEP}\
  }\textbf {\bibinfo {volume} {10}},\ \bibinfo {pages} {216} (\bibinfo {year}
  {2013})},\ \Eprint {http://arxiv.org/abs/1308.5307} {arXiv:1308.5307
  [hep-ph]} \BibitemShut {NoStop}%
\bibitem [{\citenamefont {Han}\ \emph {et~al.}(2015)\citenamefont {Han},
  \citenamefont {Kim}, \citenamefont {Munir},\ and\ \citenamefont
  {Park}}]{Han:2015lma}%
  \BibitemOpen
  \bibfield  {author} {\bibinfo {author} {\bibfnamefont {C.}~\bibnamefont
  {Han}}, \bibinfo {author} {\bibfnamefont {D.}~\bibnamefont {Kim}}, \bibinfo
  {author} {\bibfnamefont {S.}~\bibnamefont {Munir}}, \ and\ \bibinfo {author}
  {\bibfnamefont {M.}~\bibnamefont {Park}},\ }\href {\doibase
  10.1007/JHEP04(2015)132} {\bibfield  {journal} {\bibinfo  {journal} {JHEP}\
  }\textbf {\bibinfo {volume} {04}},\ \bibinfo {pages} {132} (\bibinfo {year}
  {2015})},\ \Eprint {http://arxiv.org/abs/1502.03734} {arXiv:1502.03734
  [hep-ph]} \BibitemShut {NoStop}%
\bibitem [{\citenamefont {Han}\ \emph {et~al.}(2017{\natexlab{a}})\citenamefont
  {Han}, \citenamefont {Ren}, \citenamefont {Wu}, \citenamefont {Yang},\ and\
  \citenamefont {Zhang}}]{Han:2016xet}%
  \BibitemOpen
  \bibfield  {author} {\bibinfo {author} {\bibfnamefont {C.}~\bibnamefont
  {Han}}, \bibinfo {author} {\bibfnamefont {J.}~\bibnamefont {Ren}}, \bibinfo
  {author} {\bibfnamefont {L.}~\bibnamefont {Wu}}, \bibinfo {author}
  {\bibfnamefont {J.~M.}\ \bibnamefont {Yang}}, \ and\ \bibinfo {author}
  {\bibfnamefont {M.}~\bibnamefont {Zhang}},\ }\href {\doibase
  10.1140/epjc/s10052-017-4662-7} {\bibfield  {journal} {\bibinfo  {journal}
  {Eur. Phys. J. C}\ }\textbf {\bibinfo {volume} {77}},\ \bibinfo {pages} {93}
  (\bibinfo {year} {2017}{\natexlab{a}})},\ \Eprint
  {http://arxiv.org/abs/1609.02361} {arXiv:1609.02361 [hep-ph]} \BibitemShut
  {NoStop}%
\bibitem [{\citenamefont {Han}\ \emph {et~al.}(2017{\natexlab{b}})\citenamefont
  {Han}, \citenamefont {Hikasa}, \citenamefont {Wu}, \citenamefont {Yang},\
  and\ \citenamefont {Zhang}}]{Han:2016gvr}%
  \BibitemOpen
  \bibfield  {author} {\bibinfo {author} {\bibfnamefont {C.}~\bibnamefont
  {Han}}, \bibinfo {author} {\bibfnamefont {K.-i.}\ \bibnamefont {Hikasa}},
  \bibinfo {author} {\bibfnamefont {L.}~\bibnamefont {Wu}}, \bibinfo {author}
  {\bibfnamefont {J.~M.}\ \bibnamefont {Yang}}, \ and\ \bibinfo {author}
  {\bibfnamefont {Y.}~\bibnamefont {Zhang}},\ }\href {\doibase
  10.1016/j.physletb.2017.04.026} {\bibfield  {journal} {\bibinfo  {journal}
  {Phys. Lett. B}\ }\textbf {\bibinfo {volume} {769}},\ \bibinfo {pages} {470}
  (\bibinfo {year} {2017}{\natexlab{b}})},\ \Eprint
  {http://arxiv.org/abs/1612.02296} {arXiv:1612.02296 [hep-ph]} \BibitemShut
  {NoStop}%
\bibitem [{ATL(2023)}]{ATL-PHYS-PUB-2023-005}%
  \BibitemOpen
  \href {https://cds.cern.ch/record/2852738} {\emph {\bibinfo {title} {{SUSY
  March 2023 Summary Plot Update}}}},\ \bibinfo {type} {Tech. Rep.}\ (\bibinfo
  {institution} {CERN},\ \bibinfo {address} {Geneva},\ \bibinfo {year} {2023})\
  \bibinfo {note} {all figures including auxiliary figures are available at
  https://atlas.web.cern.ch/Atlas/GROUPS/PHYSICS
  /PUBNOTES/ATL-PHYS-PUB-2023-005}\BibitemShut {NoStop}%
\bibitem [{\citenamefont {Baer}\ \emph {et~al.}(2020)\citenamefont {Baer},
  \citenamefont {Barger}, \citenamefont {Salam}, \citenamefont {Sengupta},\
  and\ \citenamefont {Sinha}}]{Baer:2020kwz}%
  \BibitemOpen
  \bibfield  {author} {\bibinfo {author} {\bibfnamefont {H.}~\bibnamefont
  {Baer}}, \bibinfo {author} {\bibfnamefont {V.}~\bibnamefont {Barger}},
  \bibinfo {author} {\bibfnamefont {S.}~\bibnamefont {Salam}}, \bibinfo
  {author} {\bibfnamefont {D.}~\bibnamefont {Sengupta}}, \ and\ \bibinfo
  {author} {\bibfnamefont {K.}~\bibnamefont {Sinha}},\ }\href {\doibase
  10.1140/epjst/e2020-000020-x} {\bibfield  {journal} {\bibinfo  {journal}
  {Eur. Phys. J. ST}\ }\textbf {\bibinfo {volume} {229}},\ \bibinfo {pages}
  {3085} (\bibinfo {year} {2020})},\ \Eprint {http://arxiv.org/abs/2002.03013}
  {arXiv:2002.03013 [hep-ph]} \BibitemShut {NoStop}%
\bibitem [{\citenamefont {Wang}\ \emph {et~al.}(2022)\citenamefont {Wang},
  \citenamefont {Wang}, \citenamefont {Yang}, \citenamefont {Zhang},\ and\
  \citenamefont {Zhu}}]{Wang:2022rfd}%
  \BibitemOpen
  \bibfield  {author} {\bibinfo {author} {\bibfnamefont {F.}~\bibnamefont
  {Wang}}, \bibinfo {author} {\bibfnamefont {W.}~\bibnamefont {Wang}}, \bibinfo
  {author} {\bibfnamefont {J.}~\bibnamefont {Yang}}, \bibinfo {author}
  {\bibfnamefont {Y.}~\bibnamefont {Zhang}}, \ and\ \bibinfo {author}
  {\bibfnamefont {B.}~\bibnamefont {Zhu}},\ }\href {\doibase
  10.3390/universe8030178} {\bibfield  {journal} {\bibinfo  {journal}
  {Universe}\ }\textbf {\bibinfo {volume} {8}},\ \bibinfo {pages} {178}
  (\bibinfo {year} {2022})},\ \Eprint {http://arxiv.org/abs/2201.00156}
  {arXiv:2201.00156 [hep-ph]} \BibitemShut {NoStop}%
\bibitem [{\citenamefont {Yang}\ \emph {et~al.}(2022)\citenamefont {Yang},
  \citenamefont {Zhu},\ and\ \citenamefont {Zhu}}]{Yang:2022qyz}%
  \BibitemOpen
  \bibfield  {author} {\bibinfo {author} {\bibfnamefont {J.~M.}\ \bibnamefont
  {Yang}}, \bibinfo {author} {\bibfnamefont {P.}~\bibnamefont {Zhu}}, \ and\
  \bibinfo {author} {\bibfnamefont {R.}~\bibnamefont {Zhu}},\ }\href {\doibase
  10.22323/1.422.0069} {\bibfield  {journal} {\bibinfo  {journal} {PoS}\
  }\textbf {\bibinfo {volume} {LHCP2022}},\ \bibinfo {pages} {069} (\bibinfo
  {year} {2022})},\ \Eprint {http://arxiv.org/abs/2211.06686} {arXiv:2211.06686
  [hep-ph]} \BibitemShut {NoStop}%
\bibitem [{\citenamefont {Ellis}\ \emph {et~al.}(2004)\citenamefont {Ellis},
  \citenamefont {Olive}, \citenamefont {Santoso},\ and\ \citenamefont
  {Spanos}}]{Ellis:2003dn}%
  \BibitemOpen
  \bibfield  {author} {\bibinfo {author} {\bibfnamefont {J.~R.}\ \bibnamefont
  {Ellis}}, \bibinfo {author} {\bibfnamefont {K.~A.}\ \bibnamefont {Olive}},
  \bibinfo {author} {\bibfnamefont {Y.}~\bibnamefont {Santoso}}, \ and\
  \bibinfo {author} {\bibfnamefont {V.~C.}\ \bibnamefont {Spanos}},\ }\href
  {\doibase 10.1016/j.physletb.2004.03.021} {\bibfield  {journal} {\bibinfo
  {journal} {Phys. Lett. B}\ }\textbf {\bibinfo {volume} {588}},\ \bibinfo
  {pages} {7} (\bibinfo {year} {2004})},\ \Eprint
  {http://arxiv.org/abs/hep-ph/0312262} {arXiv:hep-ph/0312262} \BibitemShut
  {NoStop}%
\bibitem [{\citenamefont {Covi}\ \emph {et~al.}(2001)\citenamefont {Covi},
  \citenamefont {Kim}, \citenamefont {Kim},\ and\ \citenamefont
  {Roszkowski}}]{Covi:2001nw}%
  \BibitemOpen
  \bibfield  {author} {\bibinfo {author} {\bibfnamefont {L.}~\bibnamefont
  {Covi}}, \bibinfo {author} {\bibfnamefont {H.-B.}\ \bibnamefont {Kim}},
  \bibinfo {author} {\bibfnamefont {J.~E.}\ \bibnamefont {Kim}}, \ and\
  \bibinfo {author} {\bibfnamefont {L.}~\bibnamefont {Roszkowski}},\ }\href
  {\doibase 10.1088/1126-6708/2001/05/033} {\bibfield  {journal} {\bibinfo
  {journal} {JHEP}\ }\textbf {\bibinfo {volume} {05}},\ \bibinfo {pages} {033}
  (\bibinfo {year} {2001})},\ \Eprint {http://arxiv.org/abs/hep-ph/0101009}
  {arXiv:hep-ph/0101009} \BibitemShut {NoStop}%
\bibitem [{\citenamefont {Feng}\ \emph
  {et~al.}(2003{\natexlab{a}})\citenamefont {Feng}, \citenamefont {Rajaraman},\
  and\ \citenamefont {Takayama}}]{Feng:2003uy}%
  \BibitemOpen
  \bibfield  {author} {\bibinfo {author} {\bibfnamefont {J.~L.}\ \bibnamefont
  {Feng}}, \bibinfo {author} {\bibfnamefont {A.}~\bibnamefont {Rajaraman}}, \
  and\ \bibinfo {author} {\bibfnamefont {F.}~\bibnamefont {Takayama}},\ }\href
  {\doibase 10.1103/PhysRevD.68.063504} {\bibfield  {journal} {\bibinfo
  {journal} {Phys. Rev. D}\ }\textbf {\bibinfo {volume} {68}},\ \bibinfo
  {pages} {063504} (\bibinfo {year} {2003}{\natexlab{a}})},\ \Eprint
  {http://arxiv.org/abs/hep-ph/0306024} {arXiv:hep-ph/0306024} \BibitemShut
  {NoStop}%
\bibitem [{\citenamefont {Feng}\ \emph
  {et~al.}(2003{\natexlab{b}})\citenamefont {Feng}, \citenamefont {Rajaraman},\
  and\ \citenamefont {Takayama}}]{Feng:2003xh}%
  \BibitemOpen
  \bibfield  {author} {\bibinfo {author} {\bibfnamefont {J.~L.}\ \bibnamefont
  {Feng}}, \bibinfo {author} {\bibfnamefont {A.}~\bibnamefont {Rajaraman}}, \
  and\ \bibinfo {author} {\bibfnamefont {F.}~\bibnamefont {Takayama}},\ }\href
  {\doibase 10.1103/PhysRevLett.91.011302} {\bibfield  {journal} {\bibinfo
  {journal} {Phys. Rev. Lett.}\ }\textbf {\bibinfo {volume} {91}},\ \bibinfo
  {pages} {011302} (\bibinfo {year} {2003}{\natexlab{b}})},\ \Eprint
  {http://arxiv.org/abs/hep-ph/0302215} {arXiv:hep-ph/0302215} \BibitemShut
  {NoStop}%
\bibitem [{\citenamefont {Feng}\ \emph {et~al.}(2004)\citenamefont {Feng},
  \citenamefont {Su},\ and\ \citenamefont {Takayama}}]{Feng:2004mt}%
  \BibitemOpen
  \bibfield  {author} {\bibinfo {author} {\bibfnamefont {J.~L.}\ \bibnamefont
  {Feng}}, \bibinfo {author} {\bibfnamefont {S.}~\bibnamefont {Su}}, \ and\
  \bibinfo {author} {\bibfnamefont {F.}~\bibnamefont {Takayama}},\ }\href
  {\doibase 10.1103/PhysRevD.70.075019} {\bibfield  {journal} {\bibinfo
  {journal} {Phys. Rev. D}\ }\textbf {\bibinfo {volume} {70}},\ \bibinfo
  {pages} {075019} (\bibinfo {year} {2004})},\ \Eprint
  {http://arxiv.org/abs/hep-ph/0404231} {arXiv:hep-ph/0404231} \BibitemShut
  {NoStop}%
\bibitem [{\citenamefont {Nemev\v{s}ek}\ and\ \citenamefont
  {Zhang}(2023)}]{Nemevsek:2022anh}%
  \BibitemOpen
  \bibfield  {author} {\bibinfo {author} {\bibfnamefont {M.}~\bibnamefont
  {Nemev\v{s}ek}}\ and\ \bibinfo {author} {\bibfnamefont {Y.}~\bibnamefont
  {Zhang}},\ }\href {\doibase 10.1103/PhysRevLett.130.121002} {\bibfield
  {journal} {\bibinfo  {journal} {Phys. Rev. Lett.}\ }\textbf {\bibinfo
  {volume} {130}},\ \bibinfo {pages} {121002} (\bibinfo {year} {2023})},\
  \Eprint {http://arxiv.org/abs/2206.11293} {arXiv:2206.11293 [hep-ph]}
  \BibitemShut {NoStop}%
\bibitem [{\citenamefont {Nemev\v{s}ek}\ and\ \citenamefont
  {Zhang}(2024)}]{Nemevsek:2023yjl}%
  \BibitemOpen
  \bibfield  {author} {\bibinfo {author} {\bibfnamefont {M.}~\bibnamefont
  {Nemev\v{s}ek}}\ and\ \bibinfo {author} {\bibfnamefont {Y.}~\bibnamefont
  {Zhang}},\ }\href {\doibase 10.1103/PhysRevD.109.056021} {\bibfield
  {journal} {\bibinfo  {journal} {Phys. Rev. D}\ }\textbf {\bibinfo {volume}
  {109}},\ \bibinfo {pages} {056021} (\bibinfo {year} {2024})},\ \Eprint
  {http://arxiv.org/abs/2312.00129} {arXiv:2312.00129 [hep-ph]} \BibitemShut
  {NoStop}%
\bibitem [{\citenamefont {Deshpande}\ \emph {et~al.}(2024)\citenamefont
  {Deshpande}, \citenamefont {Hamann}, \citenamefont {Sengupta}, \citenamefont
  {White}, \citenamefont {Williams},\ and\ \citenamefont
  {Wong}}]{Deshpande:2023zed}%
  \BibitemOpen
  \bibfield  {author} {\bibinfo {author} {\bibfnamefont {M.}~\bibnamefont
  {Deshpande}}, \bibinfo {author} {\bibfnamefont {J.}~\bibnamefont {Hamann}},
  \bibinfo {author} {\bibfnamefont {D.}~\bibnamefont {Sengupta}}, \bibinfo
  {author} {\bibfnamefont {M.}~\bibnamefont {White}}, \bibinfo {author}
  {\bibfnamefont {A.~G.}\ \bibnamefont {Williams}}, \ and\ \bibinfo {author}
  {\bibfnamefont {Y.~Y.~Y.}\ \bibnamefont {Wong}},\ }\href {\doibase
  10.1140/epjc/s10052-024-12992-3} {\bibfield  {journal} {\bibinfo  {journal}
  {Eur. Phys. J. C}\ }\textbf {\bibinfo {volume} {84}},\ \bibinfo {pages} {667}
  (\bibinfo {year} {2024})},\ \Eprint {http://arxiv.org/abs/2309.05709}
  {arXiv:2309.05709 [hep-ph]} \BibitemShut {NoStop}%
\bibitem [{\citenamefont {Jod\l{}owski}(2023)}]{Jodlowski:2023yne}%
  \BibitemOpen
  \bibfield  {author} {\bibinfo {author} {\bibfnamefont {K.}~\bibnamefont
  {Jod\l{}owski}},\ }\href {\doibase 10.1103/PhysRevD.108.115017} {\bibfield
  {journal} {\bibinfo  {journal} {Phys. Rev. D}\ }\textbf {\bibinfo {volume}
  {108}},\ \bibinfo {pages} {115017} (\bibinfo {year} {2023})},\ \Eprint
  {http://arxiv.org/abs/2305.05710} {arXiv:2305.05710 [hep-ph]} \BibitemShut
  {NoStop}%
\bibitem [{\citenamefont {Moroi}\ \emph {et~al.}(1993)\citenamefont {Moroi},
  \citenamefont {Murayama},\ and\ \citenamefont {Yamaguchi}}]{Moroi:1993mb}%
  \BibitemOpen
  \bibfield  {author} {\bibinfo {author} {\bibfnamefont {T.}~\bibnamefont
  {Moroi}}, \bibinfo {author} {\bibfnamefont {H.}~\bibnamefont {Murayama}}, \
  and\ \bibinfo {author} {\bibfnamefont {M.}~\bibnamefont {Yamaguchi}},\ }\href
  {\doibase 10.1016/0370-2693(93)91434-O} {\bibfield  {journal} {\bibinfo
  {journal} {Phys. Lett. B}\ }\textbf {\bibinfo {volume} {303}},\ \bibinfo
  {pages} {289} (\bibinfo {year} {1993})}\BibitemShut {NoStop}%
\bibitem [{\citenamefont {Khlopov}\ and\ \citenamefont
  {Linde}(1984)}]{Khlopov:1984pf}%
  \BibitemOpen
  \bibfield  {author} {\bibinfo {author} {\bibfnamefont {M.~Y.}\ \bibnamefont
  {Khlopov}}\ and\ \bibinfo {author} {\bibfnamefont {A.~D.}\ \bibnamefont
  {Linde}},\ }\href {\doibase 10.1016/0370-2693(84)91656-3} {\bibfield
  {journal} {\bibinfo  {journal} {Phys. Lett. B}\ }\textbf {\bibinfo {volume}
  {138}},\ \bibinfo {pages} {265} (\bibinfo {year} {1984})}\BibitemShut
  {NoStop}%
\bibitem [{\citenamefont {Hall}\ \emph {et~al.}(2010)\citenamefont {Hall},
  \citenamefont {Jedamzik}, \citenamefont {March-Russell},\ and\ \citenamefont
  {West}}]{Hall:2009bx}%
  \BibitemOpen
  \bibfield  {author} {\bibinfo {author} {\bibfnamefont {L.~J.}\ \bibnamefont
  {Hall}}, \bibinfo {author} {\bibfnamefont {K.}~\bibnamefont {Jedamzik}},
  \bibinfo {author} {\bibfnamefont {J.}~\bibnamefont {March-Russell}}, \ and\
  \bibinfo {author} {\bibfnamefont {S.~M.}\ \bibnamefont {West}},\ }\href
  {\doibase 10.1007/JHEP03(2010)080} {\bibfield  {journal} {\bibinfo  {journal}
  {JHEP}\ }\textbf {\bibinfo {volume} {03}},\ \bibinfo {pages} {080} (\bibinfo
  {year} {2010})},\ \Eprint {http://arxiv.org/abs/0911.1120} {arXiv:0911.1120
  [hep-ph]} \BibitemShut {NoStop}%
\bibitem [{\citenamefont {Eberl}\ \emph {et~al.}(2021)\citenamefont {Eberl},
  \citenamefont {Gialamas},\ and\ \citenamefont {Spanos}}]{Eberl:2020fml}%
  \BibitemOpen
  \bibfield  {author} {\bibinfo {author} {\bibfnamefont {H.}~\bibnamefont
  {Eberl}}, \bibinfo {author} {\bibfnamefont {I.~D.}\ \bibnamefont {Gialamas}},
  \ and\ \bibinfo {author} {\bibfnamefont {V.~C.}\ \bibnamefont {Spanos}},\
  }\href {\doibase 10.1103/PhysRevD.103.075025} {\bibfield  {journal} {\bibinfo
   {journal} {Phys. Rev. D}\ }\textbf {\bibinfo {volume} {103}},\ \bibinfo
  {pages} {075025} (\bibinfo {year} {2021})},\ \Eprint
  {http://arxiv.org/abs/2010.14621} {arXiv:2010.14621 [hep-ph]} \BibitemShut
  {NoStop}%
\bibitem [{\citenamefont {Cyburt}\ \emph {et~al.}(2003)\citenamefont {Cyburt},
  \citenamefont {Ellis}, \citenamefont {Fields},\ and\ \citenamefont
  {Olive}}]{Cyburt:2002uv}%
  \BibitemOpen
  \bibfield  {author} {\bibinfo {author} {\bibfnamefont {R.~H.}\ \bibnamefont
  {Cyburt}}, \bibinfo {author} {\bibfnamefont {J.~R.}\ \bibnamefont {Ellis}},
  \bibinfo {author} {\bibfnamefont {B.~D.}\ \bibnamefont {Fields}}, \ and\
  \bibinfo {author} {\bibfnamefont {K.~A.}\ \bibnamefont {Olive}},\ }\href
  {\doibase 10.1103/PhysRevD.67.103521} {\bibfield  {journal} {\bibinfo
  {journal} {Phys. Rev. D}\ }\textbf {\bibinfo {volume} {67}},\ \bibinfo
  {pages} {103521} (\bibinfo {year} {2003})},\ \Eprint
  {http://arxiv.org/abs/astro-ph/0211258} {arXiv:astro-ph/0211258} \BibitemShut
  {NoStop}%
\bibitem [{\citenamefont {Rychkov}\ and\ \citenamefont
  {Strumia}(2007)}]{Rychkov:2007uq}%
  \BibitemOpen
  \bibfield  {author} {\bibinfo {author} {\bibfnamefont {V.~S.}\ \bibnamefont
  {Rychkov}}\ and\ \bibinfo {author} {\bibfnamefont {A.}~\bibnamefont
  {Strumia}},\ }\href {\doibase 10.1103/PhysRevD.75.075011} {\bibfield
  {journal} {\bibinfo  {journal} {Phys. Rev. D}\ }\textbf {\bibinfo {volume}
  {75}},\ \bibinfo {pages} {075011} (\bibinfo {year} {2007})},\ \Eprint
  {http://arxiv.org/abs/hep-ph/0701104} {arXiv:hep-ph/0701104} \BibitemShut
  {NoStop}%
\bibitem [{\citenamefont {Holtmann}\ \emph {et~al.}(1999)\citenamefont
  {Holtmann}, \citenamefont {Kawasaki}, \citenamefont {Kohri},\ and\
  \citenamefont {Moroi}}]{Holtmann:1998gd}%
  \BibitemOpen
  \bibfield  {author} {\bibinfo {author} {\bibfnamefont {E.}~\bibnamefont
  {Holtmann}}, \bibinfo {author} {\bibfnamefont {M.}~\bibnamefont {Kawasaki}},
  \bibinfo {author} {\bibfnamefont {K.}~\bibnamefont {Kohri}}, \ and\ \bibinfo
  {author} {\bibfnamefont {T.}~\bibnamefont {Moroi}},\ }\href {\doibase
  10.1103/PhysRevD.60.023506} {\bibfield  {journal} {\bibinfo  {journal} {Phys.
  Rev. D}\ }\textbf {\bibinfo {volume} {60}},\ \bibinfo {pages} {023506}
  (\bibinfo {year} {1999})},\ \Eprint {http://arxiv.org/abs/hep-ph/9805405}
  {arXiv:hep-ph/9805405} \BibitemShut {NoStop}%
\bibitem [{\citenamefont {Aad}\ \emph {et~al.}(2023)\citenamefont {Aad} \emph
  {et~al.}}]{ATLAS:2023meo}%
  \BibitemOpen
  \bibfield  {author} {\bibinfo {author} {\bibfnamefont {G.}~\bibnamefont
  {Aad}} \emph {et~al.} (\bibinfo {collaboration} {ATLAS}),\ }\href {\doibase
  10.1103/PhysRevD.108.012012} {\bibfield  {journal} {\bibinfo  {journal}
  {Phys. Rev. D}\ }\textbf {\bibinfo {volume} {108}},\ \bibinfo {pages}
  {012012} (\bibinfo {year} {2023})},\ \Eprint
  {http://arxiv.org/abs/2304.12885} {arXiv:2304.12885 [hep-ex]} \BibitemShut
  {NoStop}%
\bibitem [{\citenamefont {Aaboud}\ \emph {et~al.}(2018)\citenamefont {Aaboud}
  \emph {et~al.}}]{ATLAS:2018nud}%
  \BibitemOpen
  \bibfield  {author} {\bibinfo {author} {\bibfnamefont {M.}~\bibnamefont
  {Aaboud}} \emph {et~al.} (\bibinfo {collaboration} {ATLAS}),\ }\href
  {\doibase 10.1103/PhysRevD.97.092006} {\bibfield  {journal} {\bibinfo
  {journal} {Phys. Rev. D}\ }\textbf {\bibinfo {volume} {97}},\ \bibinfo
  {pages} {092006} (\bibinfo {year} {2018})},\ \Eprint
  {http://arxiv.org/abs/1802.03158} {arXiv:1802.03158 [hep-ex]} \BibitemShut
  {NoStop}%
\bibitem [{\citenamefont {Lesgourgues}(2011)}]{lesgourgues2011cosmic}%
  \BibitemOpen
  \bibfield  {author} {\bibinfo {author} {\bibfnamefont {J.}~\bibnamefont
  {Lesgourgues}},\ }\href@noop {} {\bibfield  {journal} {\bibinfo  {journal}
  {arXiv preprint arXiv:1104.2932}\ } (\bibinfo {year} {2011})}\BibitemShut
  {NoStop}%
\bibitem [{\citenamefont {Blas}\ \emph {et~al.}(2011)\citenamefont {Blas},
  \citenamefont {Lesgourgues},\ and\ \citenamefont {Tram}}]{Blas:2011}%
  \BibitemOpen
  \bibfield  {author} {\bibinfo {author} {\bibfnamefont {D.}~\bibnamefont
  {Blas}}, \bibinfo {author} {\bibfnamefont {J.}~\bibnamefont {Lesgourgues}}, \
  and\ \bibinfo {author} {\bibfnamefont {T.}~\bibnamefont {Tram}},\ }\href
  {\doibase 10.1088/1475-7516/2011/07/034} {\bibfield  {journal} {\bibinfo
  {journal} {JCAP}\ }\textbf {\bibinfo {volume} {07}},\ \bibinfo {pages} {034}
  (\bibinfo {year} {2011})},\ \Eprint {http://arxiv.org/abs/1104.2933}
  {arXiv:1104.2933 [astro-ph.CO]} \BibitemShut {NoStop}%
\bibitem [{\citenamefont {Lesgourgues}\ and\ \citenamefont
  {Tram}(2011)}]{Lesgourgues:2011}%
  \BibitemOpen
  \bibfield  {author} {\bibinfo {author} {\bibfnamefont {J.}~\bibnamefont
  {Lesgourgues}}\ and\ \bibinfo {author} {\bibfnamefont {T.}~\bibnamefont
  {Tram}},\ }\href {\doibase 10.1088/1475-7516/2011/09/032} {\bibfield
  {journal} {\bibinfo  {journal} {Journal of Cosmology and Astroparticle
  Physics}\ }\textbf {\bibinfo {volume} {2011}},\ \bibinfo {pages} {032}
  (\bibinfo {year} {2011})}\BibitemShut {NoStop}%
\bibitem [{\citenamefont {Smith}\ \emph {et~al.}(2003)\citenamefont {Smith},
  \citenamefont {Peacock}, \citenamefont {Jenkins}, \citenamefont {White},
  \citenamefont {Frenk}, \citenamefont {Pearce}, \citenamefont {Thomas},
  \citenamefont {Efstathiou},\ and\ \citenamefont {Couchmann}}]{Smith:2002dz}%
  \BibitemOpen
  \bibfield  {author} {\bibinfo {author} {\bibfnamefont {R.~E.}\ \bibnamefont
  {Smith}}, \bibinfo {author} {\bibfnamefont {J.~A.}\ \bibnamefont {Peacock}},
  \bibinfo {author} {\bibfnamefont {A.}~\bibnamefont {Jenkins}}, \bibinfo
  {author} {\bibfnamefont {S.~D.~M.}\ \bibnamefont {White}}, \bibinfo {author}
  {\bibfnamefont {C.~S.}\ \bibnamefont {Frenk}}, \bibinfo {author}
  {\bibfnamefont {F.~R.}\ \bibnamefont {Pearce}}, \bibinfo {author}
  {\bibfnamefont {P.~A.}\ \bibnamefont {Thomas}}, \bibinfo {author}
  {\bibfnamefont {G.}~\bibnamefont {Efstathiou}}, \ and\ \bibinfo {author}
  {\bibfnamefont {H.~M.~P.}\ \bibnamefont {Couchmann}} (\bibinfo
  {collaboration} {VIRGO Consortium}),\ }\href {\doibase
  10.1046/j.1365-8711.2003.06503.x} {\bibfield  {journal} {\bibinfo  {journal}
  {Mon. Not. Roy. Astron. Soc.}\ }\textbf {\bibinfo {volume} {341}},\ \bibinfo
  {pages} {1311} (\bibinfo {year} {2003})},\ \Eprint
  {http://arxiv.org/abs/astro-ph/0207664} {arXiv:astro-ph/0207664} \BibitemShut
  {NoStop}%
\bibitem [{\citenamefont {Reid}\ \emph {et~al.}(2010)\citenamefont {Reid},
  \citenamefont {Percival}, \citenamefont {Eisenstein}, \citenamefont {Verde},
  \citenamefont {Spergel}, \citenamefont {Skibba}, \citenamefont {Bahcall},
  \citenamefont {Budavari}, \citenamefont {Frieman}, \citenamefont {Fukugita}
  \emph {et~al.}}]{reid2010cosmological}%
  \BibitemOpen
  \bibfield  {author} {\bibinfo {author} {\bibfnamefont {B.~A.}\ \bibnamefont
  {Reid}}, \bibinfo {author} {\bibfnamefont {W.~J.}\ \bibnamefont {Percival}},
  \bibinfo {author} {\bibfnamefont {D.~J.}\ \bibnamefont {Eisenstein}},
  \bibinfo {author} {\bibfnamefont {L.}~\bibnamefont {Verde}}, \bibinfo
  {author} {\bibfnamefont {D.~N.}\ \bibnamefont {Spergel}}, \bibinfo {author}
  {\bibfnamefont {R.~A.}\ \bibnamefont {Skibba}}, \bibinfo {author}
  {\bibfnamefont {N.~A.}\ \bibnamefont {Bahcall}}, \bibinfo {author}
  {\bibfnamefont {T.}~\bibnamefont {Budavari}}, \bibinfo {author}
  {\bibfnamefont {J.~A.}\ \bibnamefont {Frieman}}, \bibinfo {author}
  {\bibfnamefont {M.}~\bibnamefont {Fukugita}},  \emph {et~al.},\ }\href@noop
  {} {\bibfield  {journal} {\bibinfo  {journal} {Monthly Notices of the Royal
  Astronomical Society}\ }\textbf {\bibinfo {volume} {404}},\ \bibinfo {pages}
  {60} (\bibinfo {year} {2010})}\BibitemShut {NoStop}%
\bibitem [{\citenamefont {Ir\ifmmode \check{s}\else
  \v{s}\fi{}i\ifmmode~\check{c}\else \v{c}\fi{}}\ \emph
  {et~al.}(2017)\citenamefont {Ir\ifmmode \check{s}\else
  \v{s}\fi{}i\ifmmode~\check{c}\else \v{c}\fi{}}, \citenamefont {Viel},
  \citenamefont {Haehnelt}, \citenamefont {Bolton}, \citenamefont {Cristiani},
  \citenamefont {Becker}, \citenamefont {D'Odorico}, \citenamefont {Cupani},
  \citenamefont {Kim}, \citenamefont {Berg}, \citenamefont {L\'opez},
  \citenamefont {Ellison}, \citenamefont {Christensen}, \citenamefont
  {Denney},\ and\ \citenamefont {Worseck}}]{PhysRevD.96.023522}%
  \BibitemOpen
  \bibfield  {author} {\bibinfo {author} {\bibfnamefont {V.}~\bibnamefont
  {Ir\ifmmode \check{s}\else \v{s}\fi{}i\ifmmode~\check{c}\else \v{c}\fi{}}},
  \bibinfo {author} {\bibfnamefont {M.}~\bibnamefont {Viel}}, \bibinfo {author}
  {\bibfnamefont {M.~G.}\ \bibnamefont {Haehnelt}}, \bibinfo {author}
  {\bibfnamefont {J.~S.}\ \bibnamefont {Bolton}}, \bibinfo {author}
  {\bibfnamefont {S.}~\bibnamefont {Cristiani}}, \bibinfo {author}
  {\bibfnamefont {G.~D.}\ \bibnamefont {Becker}}, \bibinfo {author}
  {\bibfnamefont {V.}~\bibnamefont {D'Odorico}}, \bibinfo {author}
  {\bibfnamefont {G.}~\bibnamefont {Cupani}}, \bibinfo {author} {\bibfnamefont
  {T.-S.}\ \bibnamefont {Kim}}, \bibinfo {author} {\bibfnamefont {T.~A.~M.}\
  \bibnamefont {Berg}}, \bibinfo {author} {\bibfnamefont {S.}~\bibnamefont
  {L\'opez}}, \bibinfo {author} {\bibfnamefont {S.}~\bibnamefont {Ellison}},
  \bibinfo {author} {\bibfnamefont {L.}~\bibnamefont {Christensen}}, \bibinfo
  {author} {\bibfnamefont {K.~D.}\ \bibnamefont {Denney}}, \ and\ \bibinfo
  {author} {\bibfnamefont {G.}~\bibnamefont {Worseck}},\ }\href {\doibase
  10.1103/PhysRevD.96.023522} {\bibfield  {journal} {\bibinfo  {journal} {Phys.
  Rev. D}\ }\textbf {\bibinfo {volume} {96}},\ \bibinfo {pages} {023522}
  (\bibinfo {year} {2017})}\BibitemShut {NoStop}%
\bibitem [{\citenamefont {Boyarsky}\ \emph {et~al.}(2009)\citenamefont
  {Boyarsky}, \citenamefont {Lesgourgues}, \citenamefont {Ruchayskiy},\ and\
  \citenamefont {Viel}}]{Boyarsky:2008xj}%
  \BibitemOpen
  \bibfield  {author} {\bibinfo {author} {\bibfnamefont {A.}~\bibnamefont
  {Boyarsky}}, \bibinfo {author} {\bibfnamefont {J.}~\bibnamefont
  {Lesgourgues}}, \bibinfo {author} {\bibfnamefont {O.}~\bibnamefont
  {Ruchayskiy}}, \ and\ \bibinfo {author} {\bibfnamefont {M.}~\bibnamefont
  {Viel}},\ }\href {\doibase 10.1088/1475-7516/2009/05/012} {\bibfield
  {journal} {\bibinfo  {journal} {JCAP}\ }\textbf {\bibinfo {volume} {05}},\
  \bibinfo {pages} {012} (\bibinfo {year} {2009})},\ \Eprint
  {http://arxiv.org/abs/0812.0010} {arXiv:0812.0010 [astro-ph]} \BibitemShut
  {NoStop}%
\end{thebibliography}%

\end{document}